\documentclass[aip,jcp,twocolumn,superscriptaddress]{revtex4}%
\usepackage{epsfig,amssymb,amsmath,amsthm,amsfonts,amsbsy,mathrsfs}
\usepackage{graphicx,color}
\usepackage{amsmath}
\usepackage{amssymb}%
\usepackage{epstopdf}
\newtheorem{theorem}{Theorem}

\newtheorem{algorithm}[theorem]{Algorithm}

\newcommand{\Or}{\mathcal{O}}
\newcommand{\jump}[1]{\big[\hspace{-0.7mm} \big[ #1 \big]
  \hspace{-0.7mm} \big]}
\newcommand{\mean}[1] {\big\{ \hspace{-0.7mm} \big\{ #1 \big\}
  \hspace{-0.7mm} \big\}}
\newcommand{\abs}[1]{\left\lvert#1\right\rvert}

\newcommand{\average}[1]{\left\langle#1\right\rangle}
\newcommand{\bra}[1]{\langle#1\rvert}
\newcommand{\ket}[1]{\lvert#1\rangle}
\newcommand{\mc}[1]{\mathcal{#1}}
\newcommand{\DG}{\mathrm{DG}}
\newcommand{\eff}{\mathrm{eff}}
\newcommand{\ud}{\,\mathrm{d}}
\newcommand{\xc}{\mathrm{xc}}

\DeclareMathOperator{\Tr}{Tr}

\usepackage{listings}
\usepackage[percent]{overpic}
\usepackage{float}

\usepackage{algorithm,algorithmic}

\begin{document}

\title{Extreme-Scale Density Functional Theory High Performance Computing of DGDFT for Tens of Thousands of Atoms using Millions of Cores}

\author{Wei Hu}
\thanks{whuustc@ustc.edu.cn (Wei Hu)}
\affiliation{Hefei National Laboratory for Physical Sciences at the
Microscale, Department of Chemical Physics, and Synergetic
Innovation Center of Quantum Information and Quantum Physics,
University of Science and Technology of China, Hefei, Anhui 230026,
China}

\author{Xinming Qin}
\affiliation{Hefei National Laboratory for Physical Sciences at the
Microscale, Department of Chemical Physics, and Synergetic
Innovation Center of Quantum Information and Quantum Physics,
University of Science and Technology of China, Hefei, Anhui 230026,
China}

\author{Qingcai Jiang}
\affiliation{School of Computer Science and Technology, University
of Science and Technology of China, Hefei, Anhui 230026, China}

\author{Junshi Chen}
\affiliation{School of Computer Science and Technology, University
of Science and Technology of China, Hefei, Anhui 230026, China}

\author{Hong An}
\affiliation{School of Computer Science and Technology, University
of Science and Technology of China, Hefei, Anhui 230026, China}

\author{Weile Jia}
\affiliation{Department of Mathematics, University of California,
Berkeley, California 94720, United States}



\author{Fang Li}
\affiliation{National Supercomputing Center, Wuxi, Jiangsu 214072,
China}

\author{Xin Liu}
\affiliation{National Supercomputing Center, Wuxi, Jiangsu 214072,
China}

\author{Dexun Chen}
\affiliation{National Supercomputing Center, Wuxi, Jiangsu 214072,
China}

\author{Jinlong Yang}
\thanks{jlyang@ustc.edu.cn (Jinlong Yang)}
\affiliation{Hefei National Laboratory for Physical Sciences at the
Microscale, Department of Chemical Physics, and Synergetic
Innovation Center of Quantum Information and Quantum Physics,
University of Science and Technology of China, Hefei, Anhui 230026,
China}

\date{\today}

\pacs{ }

\begin{abstract}

High performance computing (HPC) is a powerful tool to accelerate
the Kohn-Sham density functional theory (KS-DFT) calculations on
modern heterogeneous supercomputers. Here, we describe a massively
extreme-scale parallel and portable implementation of discontinuous
Galerkin density functional theory (DGDFT) method on the Sunway
TaihuLight supercomputer. The DGDFT method uses the adaptive local
basis (ALB) functions generated on-the-fly during the
self-consistent field (SCF) iteration to solve the KS equations with
the high precision of plane-wave basis set. In particular, the DGDFT
method adopts a two-level parallelization strategy that deals with
various types of data distribution, task scheduling, and data
communication schemes, and combines with the feature of master-slave
multi-thread heterogeneous parallelism of SW26010 processor,
resulting in extreme-scale HPC KS-DFT calculations on the Sunway
TaihuLight supercomputer. We show that the DGDFT method can scale up
to 8,519,680 processing cores (131,072 core groups) on the Sunway
TaihuLight supercomputer for investigating the electronic structures
of two-dimensional (2D) metallic graphene systems that contain tens
of thousands of carbon atoms. Furthermore, we perform ab-initio
molecular dynamics simulations (AIMD) to study the process of water
adsorption on silicene by using the DGDFT method on the Sunway
TaihuLight supercomputer. We find that water molecules are
chemically dissociated on silicene at room temperature without the
need to introduce dopants or defects. Our AIMD results on water
molecules dissociation on silicene provide potential applications
for silicene-based water molecule sensors and metal-free catalysts
for the oxygen reduction reaction.

\end{abstract}

\maketitle

\section{Introduction} \label{sec:Introduction}

The Kohn-Sham density functional theory
(KS-DFT)\cite{PR_136_B864_1964_DFT, PR_140_A1133_1965_DFT} is the
most powerful methodology to perform first-principles calculations
and materials simulations for investigating the electronic
structures of molecules and solids. However, conventional KS-DFT
calculations show a cubic computational complexity
$\mathcal{O}(N^3)$ with respect to the system size $N$. The
computational cost and memory usage of the KS-DFT calculations
increases rapidly as the system size and only limited to small
systems containing hundreds of atoms. Therefore, the KS-DFT
calculations become prohibitively expensive for the first-principles
materials simulations on large-scale systems that contain thousands
of atoms.

There are several low-scaling methods have been proposed for
reducing the computational cost of KS-DFT calculations, such as
linear scaling $\mathcal{O}(N)$ methods,\cite{RMP_71_1085_1999_ON,
IRPC_29_665_2010_ON, RPP_75_036503_2010_ON} divide-and-conquer (DAC)
methods\cite{Yang1992} and fragment molecular orbital (FMO)
methods.\cite{ZhaoMezaWang2008} These low scaling methods
principally rely on the nearsightedness principle in molecules and
semiconductors and have been widely implemented with small localized
basis sets in real-space, such as
Gaussian\cite{FrischPopleBinkley1984} and numerical atomic
orbitals,\cite{IRPC_29_665_2010_ON} resulting in the sparse
Hamiltonian in real space, although they are not valid for metallic
systems. Based on these low scaling methods, several highly
efficient KS-DFT materials simulation software packages have been
developed, such as SIESTA,\cite{JPCM_14_2745_2002_SIESTA}
OPENMX,\cite{PRB_72_045121_2005_OPENMX}
CP2K,\cite{JCTC_8_3565_2012_CP2K} and
CONQUEST,\cite{CPC_177_14_2007_CONQUEST} and
HONPAS,\cite{IJQC_115_647_2014_HONPAS} which are beneficial to take
full advantage of the massive parallelism available on modern high
performance computing (HPC) architectures due to the local data
communication of the sparse Hamiltonian generated in small localized
basis sets.

However, the accuracy of these low scaling methods strongly depends
on the parameters of localized basis sets, and is difficult to be
improved systematically, compared to large uniform basis sets with
high accuracy, such as plane-waves. Several famous KS-DFT materials
simulation software have been developed using plane wave basis set,
such as VASP,\cite{PRB_47_558_1993_VASP} QUANTUM
ESPRESSO\cite{JPCM_21_395502_2009_QE} and
ABINIT.\cite{CPC_180_2580_2009_ABINIT} But such plane wave basis set
always requires large number of basis functions for the high
accuracy and is difficult to take advantage of the HPC calculations
on modern heterogeneous supercomputers due to the large all-to-all
data communications of the dense Hamiltonian.\cite{CMS_42_329_2008}

It should be noticed that massively parallel KS-DFT calculations
increasingly require more complicated software packages to achieve
better parallel performance and scalability across the vastly
diverse ecosystem of modern heterogeneous supercomputers,
especially, the widely used X86 CPU (Central Processing Unit)
architectures. In particular, large-scale KS-DFT calculations have
been performed for simulating 96,000 water molecules on 46,656 cores
in CP2K\cite{JCTC_8_3565_2012_CP2K} and 2,000,000 atoms on 4,096
cores in CONQUEST\cite{CPC_177_14_2007_CONQUEST} on the Cray
supercomputer with X86 architecture. It is well known that there are
two new generation of Chinese supercomputers, Sunway TaihuLight and
Tianhe, which are among the fastest supercomputers in the world. In
particular, the Sunway TaihuLight supercomputer use the Chinese
home-grown SW26010 processors based on a new Sunway master-slave
heterogeneous architecture different from X86 CPU architecture. Such
hardware advantage requires previous KS-DFT software to be
reoptimized and ported into the new Sunway TaihuLight supercomputer.

The recently developed discontinuous Galerkin density functional
theory (DGDFT)\cite{JCP_231_2140_2012_DGDFT,
JCP_143_124110_2015_DGDFT, PCCP_17_31397_2015_DGDFT,
JCP_145_154101_2016_CheFSI, JCP_335_426_2017_DGDFT} aims to combine
the advantages of small localized and large uniform basis sets,
which can reduce the number of basis functions similar to numerical
atomic basis sets, while maintaining the high precision comparable
to that of plane-wave basis set. The DGDFT method is discretized on
an adaptive local basis (ALB) set.\cite{JCP_231_2140_2012_DGDFT} One
unique feature of the ALB set is that each ALB function is strictly
localized in a certain element in the real space, which results in
the sparse Hamiltonian with unchanged block-tridiagonal structures
for both metallic and semiconducting systems, superior to small
localized basis sets. Therefore, the DGDFT method is beneficial to
take full advantage of the massive parallelism available on modern
heterogeneous supercomputers.\cite{JCP_143_124110_2015_DGDFT,
PCCP_17_31397_2015_DGDFT}

In the present work, we describe a massively extreme-scale parallel
and portable implementation of the DGDFT method on the Sunway
TaihuLight supercomputer. We demonstrate that the DGDFT method
adopts a two-level parallelization strategy that makes use of
different types of data distribution, task scheduling, and data
communication schemes, and combines with the feature of master-slave
multi-thread heterogeneous parallelism of SW26010 processors,
resulting in extreme-scale HPC KS-DFT calculations on the Sunway
TaihuLight supercomputer.

\section{Methodology} \label{sec:Methodology}

In this section, we describe the theoretical algorithms and parallel
implementation of the DGDFT method on the Sunway TaihuLight
supercomputer in detail. The key spirit of the DGDFT method is to
discretize the global KS equations by using the adaptive local basis
(ALB) set in discontinuous Galerkin (DG)
framework.\cite{JCP_231_2140_2012_DGDFT} That is why we call this
method as discontinuous Galerkin density functional theory
(DGDFT).\cite{JCP_143_124110_2015_DGDFT} In this work, we utilize
the Chebyshev polynomial filtered subspace iteration (CheFSI)
method\cite{JCP_145_154101_2016_CheFSI} to diagonalize the
block-tridiagonal spare DG Hamiltonian in the DGDFT method. We
present the scalable implementation of the DGDFT method based on the
two-level parallelization strategy combines with the multi-thread
parallelism and acceleration of Sunway master-slave heterogeneous
architecture, resulting in extreme-scale HPC DFT calculations on the
Sunway TaihuLight supercomputer. Finally, we use the DGDFT method to
perform first-principles KS-DFT calculations and materials
simulations for investigating the electronic structures and ab
initio molecular dynamics (AIMD) of ultra-large-scale metallic
systems containing tens of thousands
atoms.\cite{JCP_145_154101_2016_CheFSI}

\subsection{Discontinuous Galerkin density functional theory}\label{sec:DGDFT}

The basic idea of DGDFT is the domain decomposition algorithm for
generating a new type of basis sets to solve the KS
equations.\cite{JCP_231_2140_2012_DGDFT} In the DGDFT method, we
partition the global computational domain $\Omega$ into a number of
subdomains (called elements), denoted by $\mc{T} = \{E_{k}\}^M_{k =
1}$ to a collection of all elements ($M$ is the total number of
elements). In the current version of DGDFT, we use periodic boundary
conditions to treat both molecule and solids. Therefore, each
surface of the element must be shared between two neighboring
elements, and $\mc{S}$ denotes the collection of all the surfaces.

\begin{figure}[!htb]
\begin{center}
\includegraphics[width=0.5\textwidth]{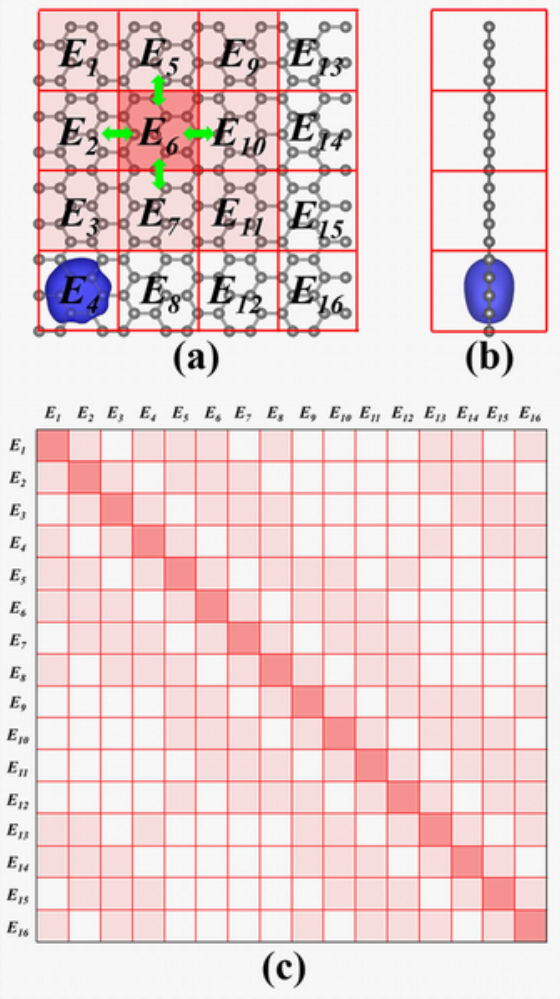}
\end{center}
\caption{A 2D graphene system G180 system in 2D partitioned into 16
(4 $\times$ 4) equal-sized elements. (a) An extended element $Q_6$
associated with the central element $E_6$, and $Q_6$ includes 9
elements in a 2D mesh ($E_1$, $E_2$, $E_3$, $E_5$, $E_6$, $E_7$,
$E_9$, $E_{10}$, and $E_{11}$). There are four surfaces surrounding
the central element $E_6$ with boundary integrals highlighted by
green arrows. The first ALB function belonging to the element $E_4$
is plotted with blue isosurfaces (0.01 Hartree/Bohr$^3$) in top and
(b) side views. (c) The block-tridiagonal spare structure of DG
Hamiltonian matrix. The blocks with nonzero values are highlighted
with red areas.} \label{fig:ALB}
\end{figure}

An example of partitioning the global domain of a graphene system
into a number of elements is given in Figure~\ref{fig:ALB}. This is
a 2D graphene system G180 containing 180 carbon atoms. The global
domain is partitioned into $16$ equal-sized elements in a 2D 4
$\times$ 4 mesh along the X and Y directions, respectively. An
extended element $Q_6$ associated with the central element $E_6$,
and $Q_6$ includes 9 elements, $E_1$, $E_2$, $E_3$, $E_5$, $E_6$,
$E_7$, $E_9$, $E_{10}$, and $E_{11}$. There are four surfaces
surrounding the central element $E_6$ with boundary integrals
highlighted by green arrows. We solve small-scale local KS equations
on this extended element $Q_6$ only containing few of atoms and
obtain a set of eigenfunctions. Then we restrict and truncate these
eigenfunctions into the central element $E_6$ and obtain a new set
of ALB functions only localized on the element $E_6$, which are
adaptive to change according to the atomic and environmental
information during the SCF iterations when solving the global KS
equations. For example, the first ALB function belonging to the
element $E_4$ is plotted in Figure~\ref{fig:ALB}(a) and (b). This
ALB function is strictly localized inside $E_{4}$ and is therefore
discontinuous across the boundary of elements, resulting in four
surfaces surround this element with boundary integrals. Therefore,
the ALB functions can be acted as a new type of localized and
orthogonal basis set to discretize the global KS equations,
resulting in a spare block-tridiagonal structure of DG Hamiltonian
matrix\cite{JCP_143_124110_2015_DGDFT} as shown in
Figure~\ref{fig:ALB}(c).

\begin{figure}[!htb]
\begin{center}
\includegraphics[width=0.5\textwidth]{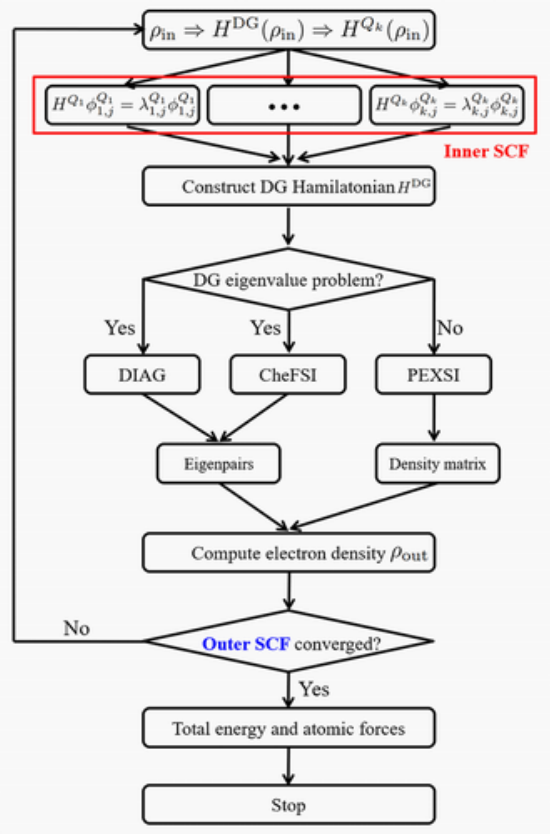}
\end{center}
\caption{Flowchart of the DGDFT method. There are four
time-consuming parts in DGDFT, including generating the ALB
functions in the inner SCF iterations, constructing and
diagonalizing the DG Hamiltonian (DIAG, CheFSI and PEXSI), computing
the electron density, total energy and atomic forces in the outer
SCF iterations. $H^{\DG}$ and $H^{Q_{k}}$ represent the global and
local $Q_{k}$ KS Hamiltonian, respectively.} \label{fig:Flowchart}
\end{figure}

It should be noticed that the framework of DGDFT to solve the global
KS equations is similar to the standard DFT methods discretized on
atomic localized basis sets, such as numerical atomic basis orbitals
implemented in the SIESTA\cite{JPCM_14_2745_2002_SIESTA} software
package. The key spirit of DGDFT is to use the ALB functions to
discretize the Kohn-Sham equations. In particular, the ALB functions
are orthogonal, localized, and complete basis set that combines with
the advantages of both numerical atomic basis orbitals
(localization) and plane waves (orthogonality and completeness).
Furthermore, such features make the DG Hamiltonian keep unchanged in
a spare block-tridiagonal structure during the SCF iterations even
for metallic systems.\cite{JCP_145_154101_2016_CheFSI} Furthermore,
there are several new diagonalization methods, such as
CheFIS\cite{JCP_145_154101_2016_CheFSI} and
PEXSI,\cite{JPCM_26_305503_2014_PEXSI} to take advantage of such
block-spare DG Hamiltonian in the framework of DGDFT. Therefore,
there are four time-consuming parts in DGDFT, including generating
the ALB functions, constructing and diagonalizing the DG
Hamiltonian, as well as computing the electron density, total energy
and atomic forces.\cite{JCP_143_124110_2015_DGDFT} For the flowchart
of DGDFT, except for the first step to generate the ALB functions
on-the-fly during the SCF iterations, other three parts of DGDFT are
similar to SIESTA. The flowchart of the DGDFT method for solving the
global KS equations is given in Figure~\ref{fig:Flowchart}.


\subsubsection{Generating ALB functions}


The key spirit of DGDFT is to generate the ALB functions on-the-fly,
which are adaptive to change according to the atomic and
environmental information during the SCF iterations when solving the
global KS equations. In order to generate the ALB functions on each
element $E_k$, we construct an extended element $Q_k$ consisting of
a central element $E_k$ and surrounding a set of buffer elements
surrounding $E_k$. An example of partitioning the global domain of a
2D graphene system is shown in the Figure~\ref{fig:ALB}(a).

On each extended element $Q_k$, we solve the local KS equations
defined as
\begin{equation}\label{eqn:LocalKS}
H^{Q_{k}} \phi^{Q_{k}}_{k,j} = \left(-\frac12 \Delta + V^{Q_{k}}_\mathrm{eff} + V^{Q_{k}}_\mathrm{nl} \right) \phi^{Q_{k}}_{k,j} = \lambda^{Q_{k}}_{k,j} \phi^{Q_{k}}_{k,j} \\
\end{equation}
where $H^{Q_{k}} $ and $\phi^{Q_{k}}_{k,j}$ are the local Humilation
and wavefunctions on the extended element $Q_k$. These local
wavefunctions $\phi^{Q_{k}}_{k,j}$ satisfy the orthonormality
condition. $V^{Q_{k}}_\mathrm{eff} = V^{Q_{k}}_\mathrm{loc} +
V^{Q_{k}}_{H} + V^{Q_{k}}_\mathrm{xc}$ is the effective potential on
the extended element $Q_k$, including the local pseudopotential
$V^{Q_{k}}_\mathrm{loc}$, the Hartree potential $V^{Q_{k}}_{H}$ and
the exchange-correlation potential $V^{Q_{k}}_\mathrm{xc}$.
$V^{Q_{k}}_\mathrm{nl} = \sum^{Q_{k}}_{I,\ell} \gamma_{I,\ell}
\ket{b_{I,\ell}} \bra{b_{I,\ell}}$ is the nonlocal pseudopotential
($\ket{b_{I,\ell}}$ is the $\ell$th projected function of the atom
$I$ and $\gamma_{I,\ell}$ is corresponding real scalar).

In the current framework of DGDFT, these local KS equations are
discretized on the standard plane-wave basis set with the same
accuracy to VASP,\cite{PRB_47_558_1993_VASP} QUANTUM
ESPRESSO\cite{JPCM_21_395502_2009_QE} and
ABINIT.\cite{CPC_180_2580_2009_ABINIT} We implement a self-contained
module called PWDFT (Plane-wave density functional theory)
\cite{JCTC_12_2242_2016_ACE, JCTC_13_5420_2017_ISDF,
JCTC_14_1311_2018_ISDF, JCTC_13_5458_2017_PCDIIS} in the DGDFT
software. When using plane wave basis set, it is viable to use the
iterative algorithms to solve the KS-DFT equations $HX = XE$.
Because the dimension of Hamiltonian matrix $H$ is above
$N^{Q_{k}}_r = 10^6$ ($N^{Q_{k}}_r$ is the number of grid points in
real space of $Q_{k}$), only a percent of amounts to more than 100
eigenvalues even for small systems containing tens of atoms. There
are several iterative algorithms have been developed to solve the
KS-DFT equations discretized on the standard plane-wave basis set,
such as Davidson\cite{Davidson1975}, Lanczos\cite{Lanczos1988},
LOBPCG,\cite{SIAMJSC_23_517_2001_LOBPCG} and
PPCG\cite{JCP_290_73_2015_PPCG} algorithms.

We compute the lowest $J_k$ eigenvalues
$\{\lambda^{Q_{k}}_{k,j}\}_{j=1}^{J_{k}}$ and corresponding
eigenfunctions $\{\phi^{Q_{k}}_{k,j}\}_{j=1}^{J_k}$ on each extended
element $Q_k$ in PWDFT. We restrict and truncate $\{\phi^{Q_{k}}_{k,
j}\}_{j=1}^{J_k}$ from the extended element $Q_{k}$ to the element
$E_k$, and obtain the truncated and orthogonal vectors
$\{\phi_{k,j}\}_{j=1}^{J_k}$ on each element $E_k$, which are the
so-called ALB functions. In the current framework of DGDFT, we set
the same number $J_k = J_b = N_b / M$ of ALB functions on each
element, where $N_b$ is the total number of ALB functions and $M$ is
the number of elements partitioned in the global domain.
Furthermore, the number of ALB functions used in each element is
$J_b \approx 4 \sim 40{ }N_e / M$, similar to the case of Gaussian
and numerical atomic basis sets. It should be noticed that the ALB
functions $\{\phi_{k,j}\}_{j=1}^{J_b}$ are truncated to zero outside
of $E_k$ and normalized orthogonally on corresponding local element
$E_k$, which lead to their discontinuity across the boundary of
$E_k$. Therefore, the ALB functions are orthogonal, localized, and
complete basis sets that combine with the advantages of both atomic
localized basis sets (localization) and plane-wave basis sets
(orthogonality, completeness and high-precision), although they are
discontinuous basis sets. By using the ALB functions to represent
the global KS equations, we call it discontinuous Galerkin density
functional theory (DGDFT).


In PWDFT, our default choice of the KS-DFT eigensolver is the LOBPCG
algorithm\cite{SIAMJSC_23_517_2001_LOBPCG} for small systems always
containing less than $20$ atoms. The LOBPCG algorithm iteratively
solves the KS-DFT eigenvalue problem of $HX = XE$ ($X =
\{x_i(r)\}_{i=1}^{J_b} \in \mathbb{R}^{N^{Q_{k}}_r \times J_b}$) by
searching the minimum of the $\text{Tr}[X^THX]$ with the
orthogonality constraint $X^TX = I \in \mathbb{R}^{J_b \times J_b}$
in a subspace spanned by $3 J_b$ vectors $[X, W, P] \in
\mathbb{R}^{N^{Q_{k}}_r \times 3 J_b}$. The eigenvectors $X$ can be
updated as
\begin{equation}
\label{eqn:4} X =  XC_{X} + WC_{W} + PC_{P}
\end{equation}
where $W$ is a preconditioned residual defined as
\begin{equation}
\label{eqn:3} W = TR = T(HX - X(X^{T}HX))
\end{equation}
where $R = HX - X(X^{T}HX)$ is the residual, and $T$ is the Teter
preconditioner widely used in the plane-wave basis set. $P$ is the
conjugate direction. The coefficients $C_{X}$, $C_{W}$ and $C_{P}$
can be computed with the lowest $J_b$ eigenpairs of the projected $3
J_b$ $\times$ $3 J_b$ generalized eigenvalue problem
\begin{equation}
\label{eqn:5} S^{T}HSC = S^{T}SCE
\end{equation}
where $S = [X, W, P]$ is the trial subspace and $C = [C_{X}$,
$C_{W}$, $C_{P}]^{T}$ are the optimal coefficients. The LOBPCG
algorithm is outlined in Algorithm~\ref{alg:LOBPCG}.


%


\begin{algorithm}[H]
\caption{The LOBPCG algorithm for iteratively solving the KS-DFT
eigenvalue problem $Hx_i = \epsilon_i x_i, i = 1,2,\dotsc,J_b$.}

\leftline{\textbf{Input:} Hamiltonian $H$ and initial wavefunctions
$\{x_i\}_{i=1}^{J_b}$.}

\leftline{\textbf{Output:} Eigenvalues
$\{\varepsilon_i\}_{i=1}^{J_b}$ and wavefunctions
$\{x_i\}_{i=1}^{N_e}$.}

\begin{algorithmic}[1]

\STATE Initialize $X$ by $\{x_i\}_{i=1}^{J_b}$ and orthonormalize
$X$.

\WHILE {convergence not reached}

\STATE Compute the preconditioned residual $W \gets T(HX -
X(X^{T}HX))$, where $T$ is the Teter preconditioner of the Laplacian
operator.

\STATE Update the trial subspace $S \gets [X, W, P]$.

\STATE Solve the projected eigenvalue problem $S^{T}HSC =
S^{T}SC\Lambda$ and obtain the coefficients $C = [C_{X}$, $C_{W}$,
$C_{P}]^{T}$.

\STATE Compute the conjugate gradient direction $P \gets WC_{W} +
PC_{P}$.

\STATE Compute $X \gets XC_{X} + P$.

\ENDWHILE

\STATE Update $\{x_i\}_{i=1}^{J_b} \gets X$.

\end{algorithmic}
\label{alg:LOBPCG}
\end{algorithm}


%

It should be noticed that when $J_b$ is relatively small ($J_b$
$\sim$ 100), the computational cost of the $3 J_b$ $\times$ $3 J_b$
projected eigenvalue problem in the Rayleigh-Ritz step in the LOBPCG
algorithm is negligible. Although the computational cost for solving
such $3 J_b$ $\times$ $3 J_b$ projected eigenvalue problem become
expensive and can no longer be ignored when $J_b$ is relatively
large ($J_b$ $\sim$ 1000-10000).

\subsubsection{Constructing DG Hamiltonian}


For the global KS equations, the wavefunctions $\psi_i(r)$ are
expanded into a linear combination of ALB functions
$\{\phi_{k,j}\}_{j=1}^{J_b}$ defined as
\begin{equation}\label{eqn:PsiALB}
  \psi_i(r) = \sum_{k=1}^{M} \sum_{j=1}^{J_b} C_{i; k, j}
  \phi_{k, j}(r)
\end{equation}

Under the ALB functions, solving the global Kohn-Sham equations
becomes a linear eigenvalue problem
\begin{equation}
  \sum_{k,j} H^{\DG}_{k', j';  k, j} C_{i; k,j} = \lambda_{i} C_{i;k',j'}
  \label{eqn:DGEigenvalue}
\end{equation}
where $H^{DG}$ is the discrete KS Hamiltonian matrix defined as
\begin{equation}
\label{eqn:DGHamiltonian}
  \begin{split}
    H^{\DG}_{k', j';  k, j} ={}&\Bigl(\frac{1}{2} \average{\nabla
    \phi_{k, j'}, \nabla \phi_{k, j}}_{\mc{T}}
    + \average{\phi_{k, j'}, V_{\eff}\phi_{k, j}}_{\mc{T}}
    \Bigr) \delta_{k,k'}\\
    &+ \Bigl(\sum_{I,\ell} \gamma_{I,\ell} \average{\phi_{k', j'},
    b_{I,\ell}}_{\mc{T}} \average{b_{I,\ell}, \phi_{k, j}}_{\mc{T}}
    \Bigr) \\
    &+ \Bigl(- \frac{1}{2} \average{\jump{\phi_{k', j'}},
    \mean{\nabla\phi_{k, j}}}_{\mc{S}} \\
    &\phantom{+ \Bigl(} - \frac{1}{2} \average{\mean{\nabla\phi_{k', j'}}, \jump{\phi_{k,
    j}}}_{\mc{S}}\\
    &\phantom{+ \Bigl(} + \alpha \average{\jump{\phi_{k', j'}}, \jump{\phi_{k,
    j}}}_{\mc{S}}\Bigr)
  \end{split}
\end{equation}
where $V_{\eff}$ denotes the effective one-body potential at the
outer SCF iterations, including local pseudopotential
$V_{\text{loc}}$, Hartree potential $V_{H}$ and the
exchange-correlation potential $V_{\xc}[\rho]$. $b_{I,\ell}$ and
$\gamma_{I,\ell}$ correspond to the nonlocal pseudopotential. For
each atom $I$, there are $L_I$ functions $b_{I,\ell}$, called the
projection vector of the nonlocal pseudopotential. The parameter
$\gamma_{I,\ell}$ is a real valued scalar. $\average{\cdot,
\cdot}_{\mc{T}}$ is the sum of the inner product on each element,
and $\average{\cdot, \cdot}_{\mc{S}}$ is the sum of the inner
product on each surface. The symbols $\mean{\cdot}$ and
$\jump{\cdot}$ represent the average and the jump operators across
surfaces respectively used to account for the discontinuity of the
basis functions.


In particular, the submatrix $H^{\DG}_{k', j'; k, j}$ is the
$(k',k)$-th matrix block of $H^{\DG} \in \mathbb{R}^{J_b \times
J_b}$. These three group terms in Eq.~\eqref{eqn:DGHamiltonian}
reflect different contributions to the DG Hamiltonian. The first
group term represents the kinetic energy and the local
pseudopotential, and only contributes to the diagonal blocks
$H^{\DG}_{k', j';  k, j}$. The second group term represents the
nonlocal pseudopotentials, and contributes to both the diagonal and
off-diagonal blocks of $H^{\DG}$. These two group terms are similar
to the case of atomic localized basis sets, such as numerical atomic
basis orbitals implemented in the SIESTA software package. However,
the third group term consists of the contributions from boundary
integrals different from SIESTA, and contributes to both the
diagonal and off-diagonal blocks of $H^{\DG}$ as well. Each boundary
term involves only two neighboring elements by definition as plotted
in Figure~\ref{fig:ALB}(a). Therefore, $H^{\DG}$ is a
block-tridiagonal spare matrix and the nonzero matrix blocks
correspond to interactions between neighboring elements as shown in
Figure~\ref{fig:ALB}(c).


\subsubsection{Diagonalizing DG Hamiltonian}


After the DG Hamiltonian is constructed by using the ALB functions,
the next step is to solve a standard eigenvalue problem to
diagonalize the DG Hamiltonian and obtain other basic physical
quantities, such as electron density, total energy and atomic
forces. Conventional method is to directly and explicitly
diagonalize the DG Hamiltonian by using the standard parallel linear
algebra software packages for dense matrices, such as the ScaLAPACK
subroutine PDSYEVD (referred as DIAG). The DIAG method is expensive
and not scalable on modern heterogeneous supercomputers, because its
computational cost scales as $\mathcal{O}(N_b^3)$ and its parallel
scalability is limited to thousands of
processors.\cite{JCP_143_124110_2015_DGDFT,
PCCP_17_31397_2015_DGDFT} In the current framework of DGDFT, we
utilize the Chebyshev polynomial filtered subspace iteration
(CheFSI)\cite{ZhouSaadTiagoEtAl2006, ZhouSaadTiagoEtAl2006a,
ZhouChelikowskySaad2014} algorithm to solve the DG eigenvalue
problem $H^{\DG} C= \Lambda C$, where $H^{\DG}$ and $C \in
\mathbb{R}^{N_b \times N_e}$.\cite{JCP_145_154101_2016_CheFSI}

The CheFSI algorithm uses a Chebyshev polynomial $p_m(\lambda)$ to
construct the map eigenvalues at the low end of occupied states
$H^{\DG}$ to the dominant eigenvalues of $p_m(H^{\DG})$. The
exponential growth property of the Chebyshev polynomials outside the
region [-1,1] can be used to obtain the wanted occupied states,
while other unwanted regions is damped in comparison.

During each CheFSI iteration, $p_m(H^{\DG}) \in \mathbb{R}^{N_b
\times N_b}$ can be applied to a block of vectors $X =
\{x_i\}^{N_e}_{i = 1} \in \mathbb{R}^{N_b \times N_e}$ by using the
three-term recurrence satisfied by Chebyshev polynomials, written as
\begin{equation}
  \label{eqn:ChebyshevHDG}
  \begin{split}
  y_{i;k, j} = & \sum_{k'=1}^{M} \sum_{j'}^{J_b} H^{\DG}_{k, j;  k', j'} x_{i;k',j'} \\
             = & \sum_{k'\in \mathcal{N}(k)} \sum_{j'}^{J_b} H^{\DG}_{k, j;  k', j'} x_{i;k',j'} \\
  \end{split}
\end{equation}
where $\mathcal{N}(k)$ denotes the collection of the neighboring
elements of the element $E_k$. These dense matrix-matrix
multiplication can be carried out independently over the various
columns of $X$, which takes advantage of the embarrassingly parallel
nature of the problem by distributing the columns among separate
processing elements.

The key step in the CheFSI algorithm is to project the DG
Hamiltonian $H^{\DG}$ onto the occupied subspace
\begin{equation}
  \label{eqn:ProjectChebyshevHDG}
  \widehat{H} = \widehat{Y}^T H^{\DG} \widehat{Y}
\end{equation}
where $\widehat{Y}$ is the orthonormal vectors for the Chebyshev
polynomial fltered block of vectors $Y= \{y_i\}^{N_e}_{i = 1} \in
\mathbb{R}^{N_b \times N_e}$. The eigenvalues $\Lambda$ and
eigenvectors $X \in \mathbb{R}^{N_b \times N_e}$ can be computed by
directly diagonalizing the projected DG Hamiltonian $\widehat{H} \in
\mathbb{R}^{N_e \times N_e}$. The pseudocode of the CheFSI algorithm
to diagonalize the DG Hamiltonian $H^{\DG}$ is shown in
Algorithm~\ref{alg:CheFSI}.
\begin{algorithm}[H]
\caption{The CheFSI algorithm for solving the global KS-DFT
eigenvalue problem $H^{\DG} X = \Lambda X, X = \{x_i\}^{N_e}_{i = 1}
\in \mathbb{R}^{N_b \times N_e}$.}

\leftline{\textbf{Input:} DG Hamiltonian $H^{\DG} \in
\mathbb{R}^{N_b \times N_b}$.}

\leftline{\textbf{Output:} Eigenvalues $\Lambda$ and eigenvectors $X
\in \mathbb{R}^{N_b \times N_e}$.}

\begin{algorithmic}[1]

\STATE Compute lower bound $b_\mathrm{low}$ by using previous Ritz
values and the upper bound $b_\mathrm{up}$ by using the Lanczos
algorithm.

\STATE Perform Chebyshev polynomial filtering $\widetilde{Y} = p_m
(H^{\DG}) X$ (m = 10 $\sim$ 40 is the filer order) with
[$b_\mathrm{low}$, $b_\mathrm{up}$] mapped to [-1; 1].

\STATE Orthonormalize columns of $\widetilde{Y}$, set $S =
\widetilde{Y}^T \widetilde{Y}$, compute $U^T U = S$, and solve
$\widehat{Y} U ^ = \widetilde{Y}$.

\STATE Diagonalize the projected DG Hamiltonian $\widehat{H} =
\widehat{Y}^T H^{\DG} \widehat{Y}$ and solve the small eigenvalue
problem $\widehat{H} V = DV$.

\STATE Perform a subspace rotation step $Y = \widehat{Y} V$.

\STATE Update $X \gets Y$.

\end{algorithmic}
\label{alg:CheFSI}
\end{algorithm}

There are three advantages for the CheFSI algorithm to take
advantage of the features of ALB functions (Orthogonality and
localization) and DGDFT framework (Block-tridiagonal spare m DG
Hamiltonian matrix). Firstly, because the ALB functions are
orthogonal, we can readily employ the orthogonal CheFSI algorithm,
which avoids to compute and orthogonalize the overlap matrix.
Secondly, because the ALB functions are localized and completed
basis sets, the number ($N_b$) of ALB functions is much smaller than
other localized basis sets, such as Gaussian and atomic numerical
basis sets. Compared to the cubic scaling O($N_b^3$) of the DIAG
method, the CheFSI method can reduce the computational cost to
O($N_b^3 N_e^2 + N_e^3$). It should be noticed that $N_b \approx 40
N_e$ for the ALB functions in DGDFT. Therefore, the CheFSI method
can speed up more than two orders of magnitude faster than the DIAG
method. Finally, orthogonal and localized ALB functions result in
the block-spare structure of the DG Hamiltonian matrix even for
metallic systems. This feature can reduce the computational cost of
the DG Hamiltonian matrix applied to a block of dense vectors with
low data communications.

The third method for diagonalizing the block-tridiagonal spare DG
Hamiltonian $H^{\DG}$ is the pole expansion and selected inversion
(PEXSI) technique.\cite{LinLuYingEtAl2009a,
JPCM_25_295501_2013_PEXSI, JPCM_26_305503_2014_PEXSI} The PEXSI
method is more efficient and scalable than the DIAG and CheFSI
methods, because the PEXSI method does not require computing
eigenvalues and eigenvectors of $H^{\DG}$. The PEXSI method is
designed for sparse matrix operations to take advantage of massively
parallel supercomputer with the high scalability, which can scale up
to 100,000 processors.\cite{JCP_143_124110_2015_DGDFT,
PCCP_17_31397_2015_DGDFT} But the PEXSI method is difficult to be
reoptimized and ported on the Sunway TaihuLight supercomputer.
Therefore, we choose the CheFSI method to diagonalize the DG
Hamiltonian in this work.


\subsubsection{Computing electron density}


After constructing the $H^{\DG}$ matrix and solving the eigenvalue
problem, the electron density can be readily evaluated from
\begin{equation}
  \label{eqn:density}
  \begin{split}
  \rho(r) = & \sum_{i=1}^{N_e} \abs{\psi_{i}(r)}^2 \\
          = & \sum_{k=1}^{M} \sum_{j=1}^{J_b}
  \sum_{k'=1}^{M} \sum_{j'=1}^{J_b} \phi_{k,j}(r)
  \phi_{k',j'}(r) \left(\sum_{i=1}^{N_e} C_{i; k,j} C_{i; k',
  j'}\right) \\
    = & \sum_{k=1}^{M} \sum_{j=1}^{J_b} \sum_{j'=1}^{J_b}
  \phi_{k,j}(r) \phi_{k,j'}(r) P_{k,j;k,j'}
  \end{split}
\end{equation}
where $P$ is the density matrix approximated as a matrix function of
$H_{DG}$ without knowing $C_{i;k,j}$ explicitly, defined as
\begin{equation}
  P_{k,j;k',j'} = \sum_{i=1}^{N_e} C_{i;k,j} C_{i;k',j'}
  \label{eqn:densitymatrix}
\end{equation}
where, each ALB function $\phi_{k,j}(x)$ is strictly localized in
the element $E_{k}$ to eliminate the cross terms involving both $k$
and $k'$. As a result, the selected blocks, or more specifically,
the diagonal blocks of the density matrix $P_{k,j;k,j'}$ are needed
to evaluate the electron density. Note that the calculation of these
blocks can be done individually on each element.

Other cheap parts, such as total
energy\cite{JCP_231_2140_2012_DGDFT} and atomic
forces,\cite{JCP_335_426_2017_DGDFT} can be rapidly computed in the
formwork of DGDFT. The total energy in the formwork of DGDFT can be
written as
\begin{equation}\label{eqn:DGvar}
  \begin{split}
    E_{\DG}(\{\psi_i\}) = {} &\frac{1}{2} \sum_{i=1}^{N_e} \average{\nabla
    \psi_i , \nabla \psi_i}_{\mc{T}}
    + \average{ V_{\eff}, \rho }_{\mc{T}}  \\
    &+ \sum_{I=1}^{N_A}\sum_{\ell=1}^{L_{I}} \gamma_{I,\ell}
    \sum_{i=1}^{N_e}
    \abs{\average{b_{I,\ell}, \psi_i}_{\mc{T}}}^2 \\
    &- \sum_{i=1}^{N_e}
    \average{\mean{\nabla\psi_i}, \jump{\psi_i}}_{\mc{S}}\\
    &+ \alpha \sum_{i=1}^{N_e} \average{\jump{\psi_i},
    \jump{\psi_i}}_{\mc{S}}
  \end{split}
\end{equation}

The atomic forces in the DGDFT method are computed with the
Hellmann-Feynman theory, and the Hellmann-Feynman force can be
compactly written as
\begin{equation}
  \begin{split}
    F_{I}^{\text{HF}} = &\int \rho_{\text{loc},I}(r-R_{I}) \nabla V_{H}(r) \ud
    r + 2 \sum_{\ell=1}^{L_{I}}\Tr[V_{\text{nl},I,\ell} P]  \\
     &+ \sum_{J\ne I}\frac{Z_{I}Z_{J}}{\abs{R_{I}-R_{J}}^3}(R_{I}-R_{J})
  \end{split}
  \label{eqn:hfforce}
\end{equation}
where $\rho_{\text{loc},I}$ is the local pseudocharge of atom $I$.
It should be noticed that the computational cost of the
Hellmann-Feynman force in DGDFT is linear scaling $\Or(N)$ with
respect to the system size.\cite{JCP_335_426_2017_DGDFT}

We have demonstrated that high plane-wave accuracy in the total
energy and atomic forces can be achieved with a very small number (4
$\sim$ 40) of basis functions per atom in the formwork of DGDFT,
compared to fully converged plane-wave
calculations.\cite{JCP_143_124110_2015_DGDFT}

\subsection{Two level parallelization strategy of DGDFT}\label{sec:Parallelization}

\begin{figure}[!htb]
\begin{center}
\includegraphics[width=0.5\textwidth]{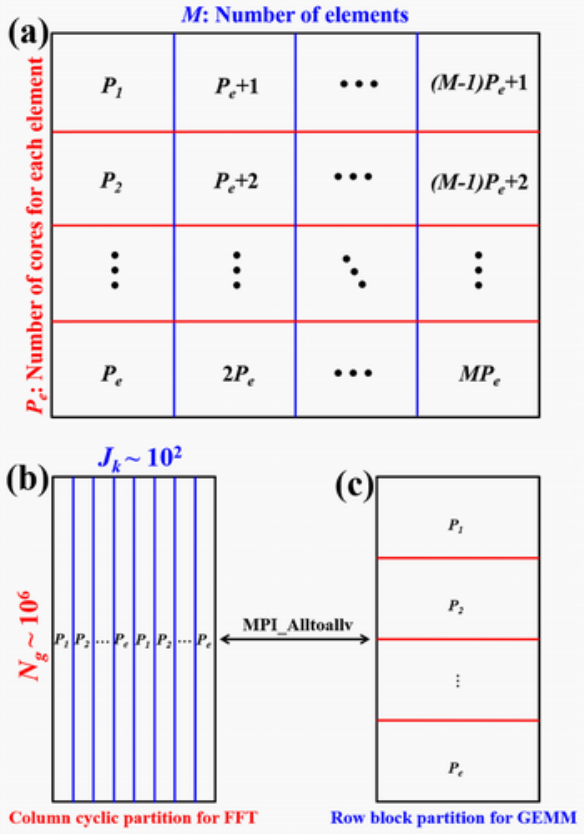}
\end{center}
\caption{(a) 2D MPI process grid for two level parallelization
strategy of DGDFT, especially for constructing and diagonalizing the
DG Hamiltonian matrix $H^{\DG} \in \mathbb{R}^{N_b \times N_b}$. $M$
is the number of elements partitioned in the system. $P_e$ is the
number of MPI processes used in each element. (b) Band
parallelization with column cyclic partition (for FFT) and (c) grid
parallelization with row block partition (for GEMM), especially for
the tall-and-skinny wavefunction matrix $\Phi^{Q_k} \in
\mathbb{R}^{N^{Q_k}_r \times J_k}$ ($J_k = J_b \approx 4 \sim 40{
}N_e / M$) on the extended element $Q_k$ in PWDFT.}
\label{fig:Parallelization}
\end{figure}

In the framework of DGDFT, there is two-level of parallelization
that deals with different types of data distribution and
communication, and task scheduling schemes as shown in
Figure~\ref{fig:Parallelization}, resulting in extreme-scale HPC
parallelism on modern heterogeneous supercomputers. The DGDFT method
use the Message Passing Interface (MPI) for parallel programming to
deal with the data communications between different MPI processes.

The first main level is called inter-element parallelization between
neighboring elements. The main computation of this level is to
construct the DG Hamiltonian matrix by using the ALB functions as
shown in Figure~\ref{fig:Parallelization}(a), which uses the column
block MPI grid partition for the global 2D MPI process grid. Because
the ALB functions are orthogonal and localized basis sets, the DG
Hamiltonian matrix can keep in a sparse block-tridiagonal structure
unchanged during the outer SCF iterations even for metallic systems.
The local data communication between neighboring elements is dealt
with the MPI programming. In the current framework of DGDFT, the
number of MPI processes can be used in this level which is fixed and
equal to the number $M$ of elements partitioned in the system. It
should be noticed that this level of inter-element parallelization
in DGDFT is highly scalable due to the local data communication
between neighboring elements for constructing the block-tridiagonal
DG Hamiltonian matrix on tens of thousands of cores on modern
heterogeneous HPC supercomputers.

The secondary level is called intra-element parallelization on each
element. This level parallelization uses the row block MPI grid
partition for the global 2D MPI process grid. The generation of ALB
functions for solving the local Kohn-Sham equations on each element
in the inner SCF iterations can be efficiently parallelized similar
to the case of conventional standard plane-wave DFT software
packages, such as VASP,\cite{PRB_47_558_1993_VASP} QUANTUM
ESPRESSO\cite{JPCM_21_395502_2009_QE} and
ABINIT.\cite{CPC_180_2580_2009_ABINIT} The DGDFT software includes a
self-contained module called PWDFT\cite{JCTC_12_2242_2016_ACE} for
performing conventional standard plane-wave based electronic
structure calculations. It should be noticed that plane-wave basis
sets always require relatively large number of basis functions for
high-accuracy KS-DFT calculations. There are two types of
parallelization, called the band and grid parallelization
(Figure~\ref{fig:Parallelization}(b) and (c)),\cite{CMS_42_329_2008}
in the inner SCF iterations of PWDFT for each row block MPI process
$P_e$ in the global 2D MPI process grid. In particular, for the
tall-and-skinny wavefunction matrix $\Phi^{Q_k} \in
\mathbb{R}^{N^{Q_k}_r \times J_b}$ on the extended element $Q_k$ in
PWDFT, the band parallelization is to use the column cyclic
partition especially for matrix operations of FFTs, while the grid
parallelization is to use row block partition especially for matrix
operations of matrix multiplications of GEMMs. We use the
MPI\_Alltoallv function to transfer two types of data partition
between the band and grid parallelization. However, large basis sets
are not conducive to take full advantage of HPC on modern
heterogeneous supercomputers due to the high all-to-all data
communication of the dense Hamiltonian matrix generated in large
uniform plane-wave basis sets. Therefore, PWDFT can only deal with
small-scale systems containing hundreds of atoms and scale to
thousands of cores.\cite{JCTC_12_2242_2016_ACE} Because each element
only contains less than tens of atoms in DGDFT, in the intra-element
parallelization, we perform the small-scale KS-DFT calculations for
solving the local KS equations by using $P_e < $ 200 cores on each
element with ultrahigh parallel efficiency of 95\% in the inner SCF
iterations of PWDFT. In the current framework of DGDFT, the maximum
number can be used this level for PWDFT is $J_b = N_b / M$, thus, 1
$\leq P_e \leq N_b / M$. It should be noticed that this level of
intra-element parallelization in DGDFT only requires to scale to
hundreds of cores for such small-scale KS-DFT calculations on modern
heterogeneous HPC supercomputers.

For the global outer SCF iterations, diagonalizing the DG
Hamiltonian matrix is the most expensive part of DGDFT for
large-scale materials simulations, three diagonalization methods
(DIAG, CheFSI and PEXSI)\cite{JCP_143_124110_2015_DGDFT,
JCP_145_154101_2016_CheFSI} can directly take advantage of such
two-level parallelization strategy of 2D MPI process grid in the
DGDFT method. From such two-level parallelization strategy, the
minimum number $N_\mathrm{min}$ and maximum number $N_\mathrm{max}$
of MPI processes used in DGDFT are computed as $N_\mathrm{min} = M$
and $N_\mathrm{max} = MJ_b = N_b$. By using this two-level
parallelization strategy, DGDFT is highly scalable on hundreds of
thousands of cores on modern heterogeneous HPC
supercomputers.\cite{JCP_143_124110_2015_DGDFT}

\subsection{DGDFT on Sunway TaihuLight supercomputer}\label{sec:Sunway}

\begin{figure}[!htb]
\begin{center}
\includegraphics[width=0.5\textwidth]{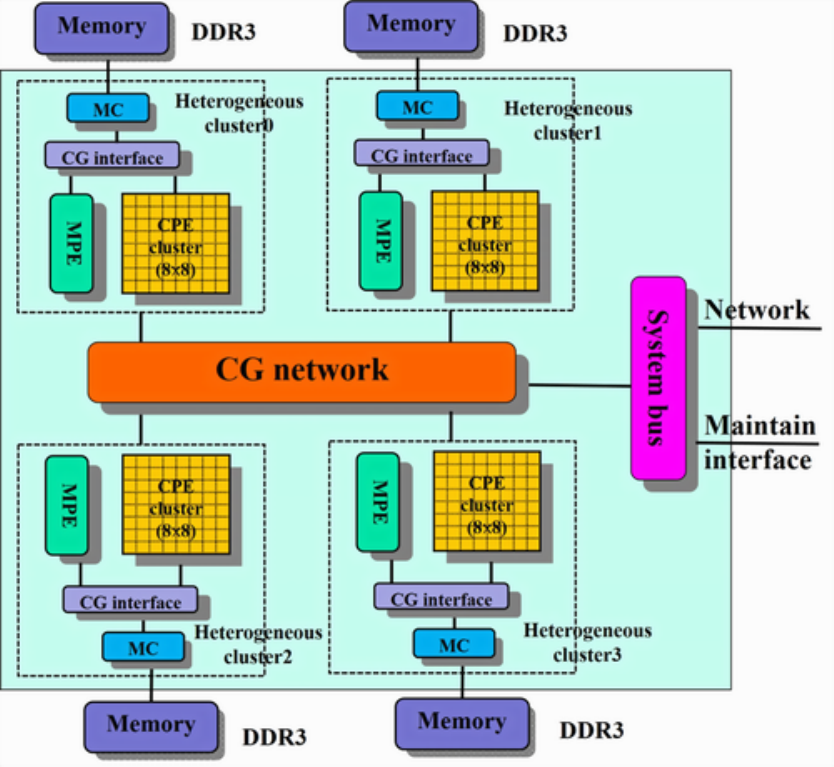}
\end{center}
\caption{The SW26010 processor architecture in the Sunway TaihuLight
supercomputer.} \label{fig:SunwayChip}
\end{figure}

The Sunway TaihuLight is the new generation of Chinese home-grown
supercomputer, and it ranks the No. 3 on the top500 list in 2019. It
consists of 40960 domestic-designed SW26010 processors, which is
based on a master-slave heterogeneous architecture. It should be
noted that Sunway processor uses a Reduced Instruction Set Computer
(RISC) design, different from the widely used X86 Complex
Instruction Set Computer (CISC) architecture. The architecture of
the SW26010 processor is shown in Figure~\ref{fig:SunwayChip}. Each
SW26010 processor chip contains 4 core groups (CGs), and each CG
acts as a master-slave many-core module. In a single CG module, one
management processing element (MPE) works as the master core and 64
computing processing elements (CPEs) arranged in an 8 $\times$ 8
grid serve as the slave cores.


We have implemented the DGDFT software package in the C/C++
programming language and uses the message passing interface (MPI)
for parallel programming. For DGDFT, each MPI process runs on a CG
with MPE as a master processing core and can be effectively
multi-thread accelerated by 64 PCEs as slave processing cores on the
SW26010 processor, similar to the graphics processing unit (GPU)
parallel programming.

In particular, the two-level parallelization strategy of DGDFT acts
as a process-level parallelism between the MPEs of CGs and the slave
processing acceleration can be considered as a thread-level
parallelism by using the 64 slave processing cores on each CG on the
Sunway TaihuLight supercomputer. Therefore, the minimum and maximum
numbers of processing cores used in DGDFT for the KS-DFT
calculations respectively are 65$M$ and 65$N_b$, where $M$ is the
number of elements and $N_b$ is the number of the ALB functions used
in the system.

The time-consuming cost in DGDFT is spent in the matrix operations,
such as the vector-vector, matrix-vector, and matrix-matrix
multiplications (DGEMM), matrix diagonalization (DSYEVD), matrix
Cholesky factorization (DPOTRF) and fast Fourier transform (FFT).
All of these matrix operations can be realized though the Basic
Linear Algebra Subprograms (BLAS), Linear Algebra PACKage (LAPACK)
and Fastest Fourier Transform in the West (FFTW) libraries.
Fortunately, most of these subroutines in the BLAS and LAPACK
libraries have been well optimized and accelerated by the slave
processing CPEs on each CG of SW26010 processors through the
thread-level parallelism on the Sunway TaihuLight supercomputer.

For example, for each CG with 65 processing cores (1 MPE and 64
PCEs), the non-transposed and transposed matrix multiplication DGEMM
subroutines in BLAS can be accelerated by factors of 3.99 and 51.53,
respectively. The DSYEVD subroutine in LAPACK can been accelerated
by a factor of 8.30. It should be noticed that their corresponding
parallel libraries, PBLAS and ScaLAPACK, can be also accelerated
automatically at the same time, similar to the cases of BLAS and
LAPACK. Unfortunately, the real-to-complex FFTs used in DGDFT have
not been optimized yet on the Sunway TaihuLight supercomputer.

\section{Results and discussion} \label{sec:Results}

In this section, we demonstrate the computational efficiency and
parallel scalability of the DGDFT method to accelerate large-scale
KS-DFT calculations on the Sunway TaihuLight supercomputer. We have
implemented the DGDFT method as software package also called
DGDFT,\cite{JCP_143_124110_2015_DGDFT} which has been written in the
C/C++ programming language with the message passing interface (MPI)
for parallel programming. The DGDFT software supports the
Hartwigsen-Goedecker-Hutter (HGH)\cite{PRB_58_3641_1998_HGH}
norm-conserving pseudo-potential. In this work, we use the
exchange-correlation functional of local density approximation of
Goedecker-Teter-Hutter (LDA-PZ)\cite{PRB_54_1703_1996_LDA} to
describe the electronic structures of metallic graphene systems. It
should be noticed that the computational accuracy of the DGDFT
method is comparable to standard plane-wave KS-DFT calculations,
such as QUANTUM ESPRESSO\cite{JPCM_21_395502_2009_QE} and
ABINIT,\cite{CPC_180_2580_2009_ABINIT} which has already been
validated in our previous works.\cite{JCP_231_2140_2012_DGDFT,
JCP_143_124110_2015_DGDFT, PCCP_17_31397_2015_DGDFT,
JCP_145_154101_2016_CheFSI, JCP_335_426_2017_DGDFT} In this work, we
focus on its computational efficiency and parallel scalability of
the DGDFT method accelerated by the extreme-scale HPC master-slave
multi-thread heterogeneous parallelism on the Sunway TaihuLight
supercomputer.

\begin{figure}[!htb]
\begin{center}
\includegraphics[width=0.5\textwidth]{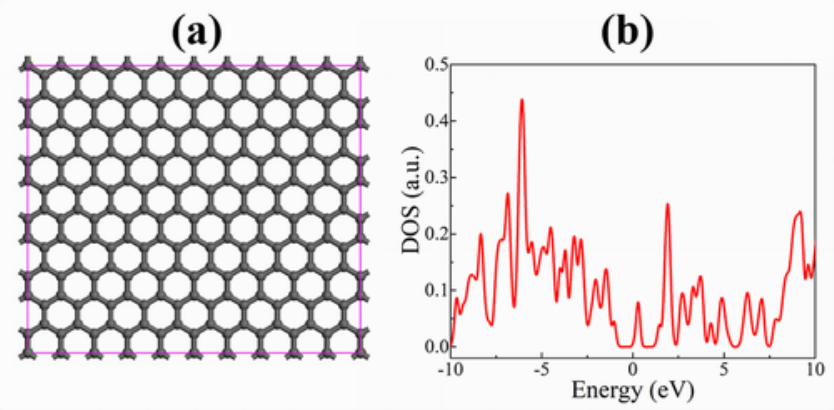}
\end{center}
\caption{(a) Geometric structures and (b) and total density of
states (DOS) of metallic graphene system G180. The gray balls denote
carbon atoms. The Fermi level is set to zero.} \label{fig:Graphene}
\end{figure}

We use the DGDFT method to investigate the electronic structures of
three graphene systems, G180, G2880 and G11520, containing 180, 2880
and 11520 carbon atoms, respectively. The geometric structures and
total densities of states (DOS) of G180 are computed and plotted in
Figure~\ref{fig:Graphene}. The DOS calculations show that the G180
system is metallic. The G2880 and G11520 systems are generated from
extending the G180 system in the $x$ and $y$ directions with 4
$\times$ 4 and 8 $\times$ 8 supercells, respectively. All these
systems are closed shell systems, and the number of occupied states
is $N_o$ = $N_e$/2.

In order to check the acceleration performance of the master-slave
multi-thread heterogeneous parallelism of SW26010 processors on the
Sunway TaihuLight supercomputer, we set a high kinetic energy cutoff
$E_\mathrm{cut}$ for three metallic graphene systems
($E_\mathrm{cut}$ = 65 Ha for G180 and G2880, and $E_\mathrm{cut}$ =
55 Ha for G11520). We set 1 $\sim$ 10 times inner SCF iterations in
PWDFT to generate the ALB functions with high accuracy for metallic
graphene systems.

Table~\ref{table:Parameters} lists the computational parameters of
graphene (G180, G2880 and G11520) systems used in the DGDFT method,
including the number $N_r$ of grid points, the numbers $N_A$ of
carbon atoms, the numbers $N_e$ of electrons,  the number $J_b$ of
the ALB functions used in each element, the numbers $M$ of elements,
the numbers $N_b$ of the ALB functions, and the minimum
$N_\mathrm{min} = M$ and maximum $N_\mathrm{max} = MJ_b = N_b$
numbers of CGs used on the Sunway TaihuLight supercomputer. The
total number of grid points $N_g$ in real space is determined from
the kinetic energy cutoff $E_\mathrm{cut}$ defined as $(N_r)_i =
\sqrt{2E_\mathrm{cut}}L_i / \pi$, where $L_i$ is the length of
supercells along the $i$-th ($x$, $y$ and $z$) coordinate direction.
\begin{table}[!tb]
\centering \caption{Computational parameters of graphene (G180,
G2880 and G11520) systems in DGDFT, including the numbers $N_A$ of
carbon atoms, the numbers $N_e$ of electrons, the number $N_r$ of
grid points in real space, the number $J_b$ of the ALB functions
used in each element, the numbers $M$ of elements, the numbers $N_b$
of the ALB functions, and the minimum ($N_\mathrm{min} = M$) and
maximum ($N_\mathrm{max} = MJ_b = N_b$) numbers of MPI processes
(CGs) used on the Sunway TaihuLight supercomputer.}
\label{table:Parameters}
\begin{tabular}{ccccccc} \ \\
\hline \hline
Systems & $N_A$  & $N_e$  & $N_r$        & $J_b$ &  $N_\mathrm{min} = M$   & $N_\mathrm{max} = N_b$ \ \\
\hline
G180    & 180    & 720    & 633,600      &  200  &  16    &   3,200   \ \\
G2880   & 2,880  & 11,520 & 183,500,800  &  128  & 1,024  & 131,072   \ \\
G11520  & 11,520 & 46,080 & 552,075,264  &  100  & 2,304  & 230,400   \ \\
\hline \hline
\end{tabular}
\end{table}

\begin{figure}[!htb]
\begin{center}
\includegraphics[width=0.5\textwidth]{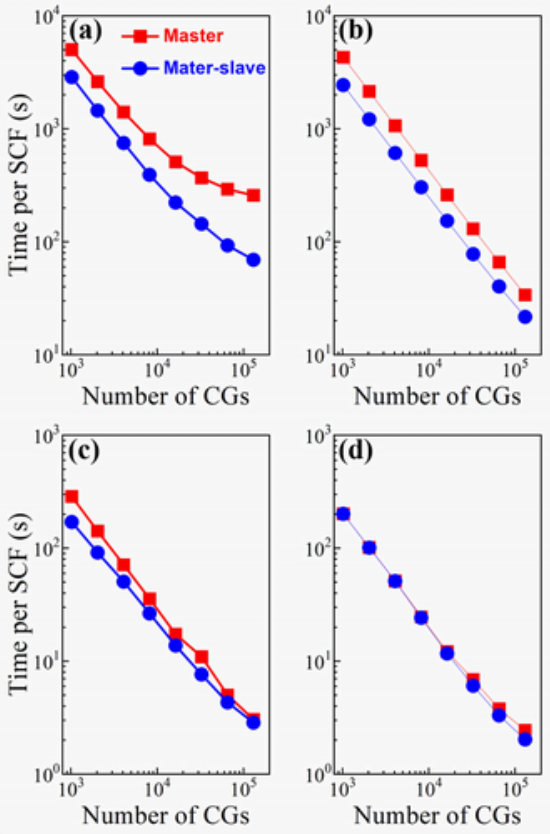}
\end{center}
\caption{The wall clock time with respect to the number of CGs
without or with the master-slave multi-thread parallelism on the
Sunway TaihuLight supercomputer used for 2D metallic graphene system
Si2880 containing 2880 carbon atoms. Strong scaling of (a) total
computational time per outer SCF iteration, including the time for
(b) generating the ALB functions, (c) constructing the DG
Hamiltonian, and (d) computing the electron density.}
\label{fig:TimeSi2880}
\end{figure}

To illustrate the computational efficiency and parallel scalability
of the DGDFT method, we demonstrate the computational time of four
time-consuming parts for 2D metallic graphene system (G2880 and
G11520) without or with the master-slave multi-thread parallelism on
the Sunway TaihuLight supercomputer, including generating the ALB
functions, constructing and diagonalizing the DG Hamiltonian, and
computing the electron density, as shown in
Figure~\ref{fig:TimeSi2880} and Figure~\ref{fig:TimeG11520}. There
are some additional steps such as computing total energy and atomic
forces, which are all included in the total wall clock time of outer
SCF iterations in the DGDFT method.

\subsection{Master process parallelism} \label{sec:Master}

We first validate the parallel scalability of DGDFT when only using
the master process parallelism on the CGs of Sunway architecture but
without slave multi-thread acceleration for the G2880 system. In
this case, each CG acts as a MPI process similar to a core in the
CPU processor of the widely used X86 architecture. For the G2880
system, we set a large number (45.5) of ALB functions per atom
($J_b$ = 128, $M$ = 1,024 and $N_b$ = 131,072) and 5 times inner SCF
iterations to generate the ALB functions for achieving high
accuracy. In this case, generating the ALB functions becomes the
most expensive part in the KS-DFT calculations by using the DGDFT
method.

The total time of the DGDFT calculations for the G2880 system is
5024.35 and 257.45 s by using $N_\mathrm{min}$ = 1,024 and
$N_\mathrm{max}$ = 131,072 CGs, respectively. The parallel
efficiency is only 15.24\% (The speedup is 19.51) when using 131,072
master CGs. In detail, when using 1,024 master CGs, the time for
four expensive parts are 4296.46 s for generating the ALB functions,
227.19 and 201.33 s for constructing and diagonalizing the DG
Hamiltonian, and 201.33 s for computing the electron density,
respectively. The time of generating the ALB functions, constructing
the DG Hamiltonian, and computing the electron density is reduced to
33.79, 3.05 and 2.44 s, respectively, when using 131,072 master CGs.
And corresponding parallel efficiencies can achieve as high as
99.33\%, 58.19\%, and 64.46\%, respectively.

The main part of reducing the total parallel efficiency is to
diagonalize the DG Hamiltonian by using the CheFSI method in DGDFT.
It should be noticed that the time of diagonalizing the DG
Hamiltonian is only reduced from 227.19 to 210.61 s when increasing
the number of master CGs from 1,024 to 131,072. The major bottleneck
of the CheFSI method is to solve the projected subspace eigenvalue
problem, which can only use the column block MPI grid partition
(1,024 MPI processes) in the global 2D MPI process grid. The time
for solving this small eigenvalue problem is almost unchanged and
dominated in the CheFSI method for the G2880 system. Other parts in
the CheFSI method only require the block-matrix multiplications,
which can ideally take advantage of the global 2D MPI process grid.
Therefore, the computation of these parts is highly scalable
parallelized by using the master process parallelism of CGs and the
time is negligible in this case for the G2880 system.

\subsection{Master-slave multi-thread parallelism} \label{sec:Slave}

The major advantage of Chinese home-grown SW26010 processors is
based on a new Sunway master-slave heterogeneous architecture, which
can be efficiently accelerated by the master-slave multi-thread
parallelism on the Sunway TaihuLight supercomputer.

In this case, when using the master-slave multi-thread parallelism
for the G2880 system, the total time can be further reduced to
2863.05 and 69.17 s by using $N_\mathrm{min}$ = 1,024 and
$N_\mathrm{max}$ = 131,072 CGs, respectively, which is much faster
than the case (5024.35 and 257.45 s) of the master process
parallelism. Furthermore, the total parallel efficiency increases to
32.33\% (The speedup is 41.39) when using 131,072 CGs with the
master-slave multi-thread parallelism.

In detail, when using 1,024 CGs (66560 processing cores), the time
for four expensive parts are 2434.22 s for generating the ALB
functions, 170.54 and 36.90 s for constructing and diagonalizing the
DG Hamiltonian, and 200.89 s for computing the electron density,
respectively. In particular, three parts of generating the ALB
functions, constructing and diagonalizing the DG Hamiltonian have
been accelerated by factors of 1.76, 1.33 and 5.45 by using the
master-slave multi-thread parallelism.

But the part of computing the electron density has not been
accelerated by using such master-slave multi-thread parallelism,
that is because that most matrix operations of this part are
real-to-complex FFTs and MPI data commutations between CGs, which
can not benefit from the master-slave multi-thread parallelism on
the Sunway TaihuLight supercomputer.

\subsection{Tens of thousands of atoms materials simulations} \label{sec:G11520}

As we know that it is a major challenge to perform the KS-DFT
calculations for first-principles materials simulations on
ultra-large-scale systems containing tens of thousands of atoms,
especially for metallic systems. The computational cost and memory
usage of such ultra-large-scale KS-DFT calculations becomes
prohibitively expensive for tens of thousands of atoms materials
simulations. We demonstrate that DGDFT can be used to push the
envelope to investigate the electronic structures of
ultra-large-scale metallic systems containing tens of thousands of
atoms by combing with the theoretical algorithms and two-level
parallelization strategy of DGDFT, and the master-slave multi-thread
heterogeneous parallelism of the Sunway TaihuLight supercomputer.

\begin{figure}[!htb]
\begin{center}
\includegraphics[width=0.5\textwidth]{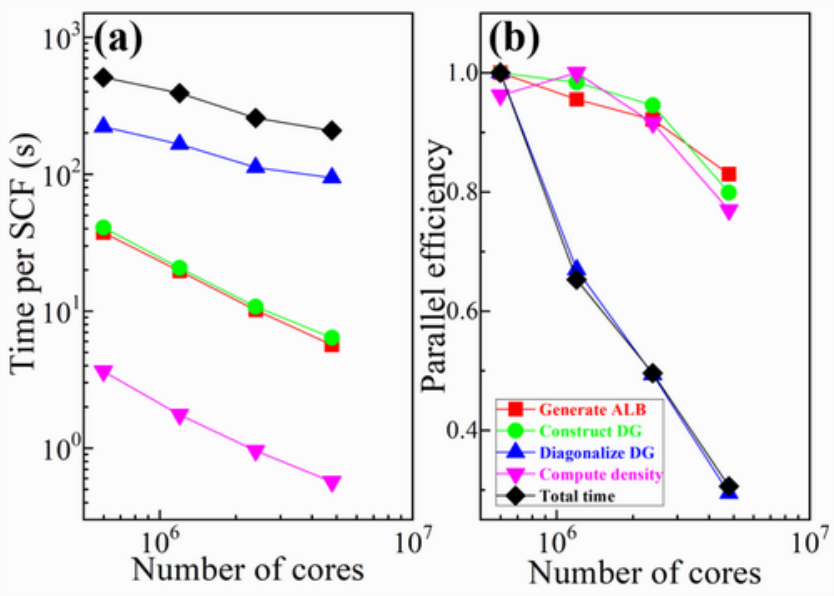}
\end{center}
\caption{The wall clock time with respect to the number of cores by
using the master-slave multi-thread parallelism on the Sunway
TaihuLight supercomputer used for 2D metallic graphene system G11520
containing 11520 carbon atoms. (a) Strong scaling and (b) parallel
efficiency of total computational time per SCF iteration, including
the time for generating the ALB functions, constructing and
diagonalizing the DG Hamiltonian, and computing the electron
density.} \label{fig:TimeG11520}
\end{figure}

Figure~\ref{fig:TimeG11520} shows the wall clock time with respect
to the number of cores by using the master-slave multi-thread
parallelism on the Sunway TaihuLight supercomputer for the G11520
system containing 11520 carbon atoms. Because these KS-DFT
calculations are ultra-large-scale materials simulations, we only
set a small number (20.0) of ALB functions per atom ($J_b$ = 100,
$M$ = 2,304 and $N_b$ = 230,400) and once inner SCF iteration to
generate the ALB functions for the G11520 system. In this case,
diagonalizing the DG Hamiltonian becomes the most expensive part in
such ultra-large-scale KS-DFT calculations by using the DGDFT
method. Other three parts become much cheaper and more scalable than
that of diagonalizing the DG Hamiltonian.

The total time of the DGDFT calculations for the G11520 system is
509.90 and 208.2 s when respectively using 9,216 and 73,728 CGs
(599,040 and 4,792,320 cores) with the master-slave multi-thread
parallelism. The parallel efficiency is 30.61\% (The speedup is
2.45) when using 73,728 CGs (4,792,320 cores). In detail, when using
599,040 cores, the time for four expensive parts is 37.51 s for
generating the ALB functions, 40.80 and 221.75 s for constructing
and diagonalizing the DG Hamiltonian, and 3.63 s for computing the
electron density, respectively. These time is reduced to 5.65, 1.38,
94.16, and 0.56 s, respectively, when using 4,792,320 cores. It
should be noticed that three cheap parts, including generating the
ALB functions, constructing the DG Hamiltonian, and computing the
electron density, show high parallel efficiencies up to 82.99\%,
79.90\%, and 76.95\%, respectively, when using 4,792,320 cores. But
the parallel efficiency of the expensive part to diagonalize the DG
Hamiltonian by using the CheFSI method is only 29.43\% when using
4,792,320 cores.


\subsection{Applications to water dissociation on silicene} \label{sec:Applications}

It is well known that 2D graphene has a much larger reactive contact
area for water adsorption\cite{NatureMater_6_652_2007} than
conventional 3D metals and metal oxides, making it ideal for use as
a surface catalyst for water chemisorption and
dissociation.\cite{ChemRev_106_1478_2006} However, water molecules
are physically adsorbed on graphene with small adsorption energies
via weak van der Waals interactions.\cite{PRB_77_125416_2008,
PRB_79_235440_2009, PRB_85_085425_2012} Fortunately, silicene, as a
2D silicon monolayer analogous to graphene but with a buckled
honeycomb structure, exhibits a much higher chemical reactivity than
graphene for water adsorption.\cite{NanoResearch_10_2223_2017}

\begin{figure}[!htb]
\begin{center}
\includegraphics[width=0.5\textwidth]{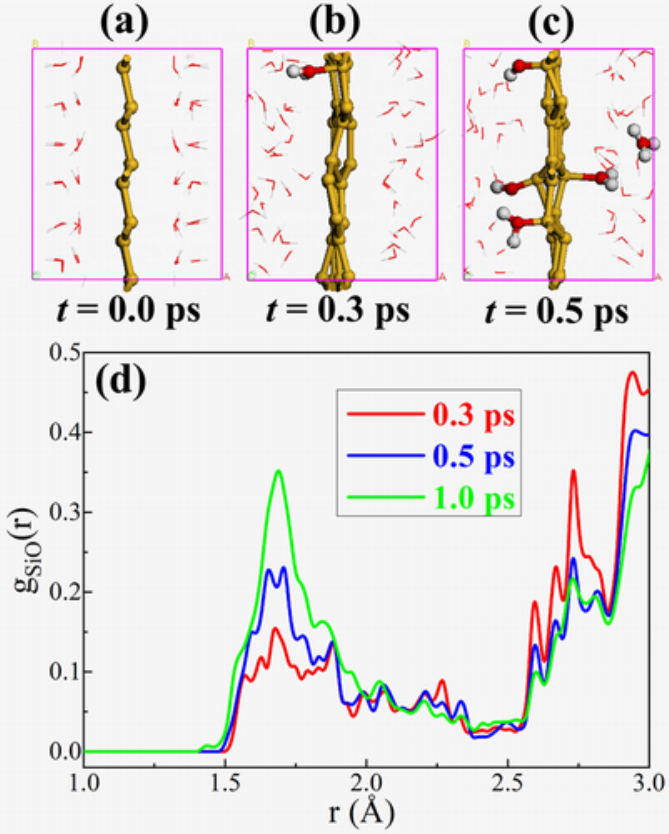}
\end{center}
\caption{AIMD simulations of liquid water adsorption on silicene.
Three snapshots of liquid water adsorption on silicene
Si$_{32}$H$_{96}$O$_{48}$ are shown at the time of (a) t = 0.0 ps,
(b) t = 0.3 ps, (c) t = 0.5 ps. Water molecules chemically adsorbed
and dissociated on silicene are highlighted. The yellow, red, and
white balls denote silicon, oxygen, and hydrogen atoms,
respectively. (d) The silicon-oxygen radial distribution function
g$_{SiO}$(r) of liquid water adsorption on silicene is computed from
AIMD simulations at T = 300 K.} \label{fig:H2OSilicene}
\end{figure}

We perform ab-initio molecular dynamics simulations (AIMD) to study
the process of water adsorption on silicene by using the DGDFT
method on the Sunway TaihuLight supercomputer. We use the
generalized gradient approximation of Perdew-Burke-Ernzerhof
(GGA-PBE)\cite{PRL_77_3865_1996_PBE} exchange-correlation functional
with semi-empirical long-range dispersion correction proposed by
Grimme (DFT-D2)\cite{JCC_27_1787_2006_Grimme} to describe the
electronic structures of water adsorption on silicene. The AIMD
simulations are performed for about 1.0 ps with a time step of 1.0
fs at a temperature of 300 K, controlled by a Nose-Hoover
thermostat.\cite{PRA_31_1695_1985_Hoover} The AIMD simulations are
used to study three systems, Si$_{32}$H$_{96}$O$_{48}$
(Figure~\ref{fig:H2OSilicene}), Si$_{128}$H$_{384}$O$_{192}$, and
Si$_{512}$H$_{1536}$O$_{768}$, including 176, 704, and 2816 atoms,
respectively. The density of liquid water (H$_{96}$O$_{48}$,
H$_{384}$O$_{192}$, and H$_{1536}$O$_{768}$) is about 1.0
g$\cdot$cm$^{-3}$.

We find that the AIMD simulations of these three systems give
similar results. The AIMD simulations of liquid water adsorption on
silicene for the Si$_{32}$H$_{96}$O$_{48}$ system are shown in
Figure~\ref{fig:H2OSilicene}. In the initial configuration (t = 0.0
ps), water molecules are set to be physically adsorbed on silicene.
At t = 0.3 ps, one water molecule is chemically adsorbed on
silicene. At t = 0.5 ps, more water molecules are chemically
adsorbed, and some have even dissociated on silicene at room
temperature. We find that more water molecules have become
chemically adsorbed and dissociated on silicene at room temperature
after t = 1.0 ps.

To further confirm that water molecules are chemically adsorbed and
dissociated on silicene at room temperature (300 K), the
silicon-oxygen radial distribution function of liquid water
adsorption on silicene is computed from the AIMD snapshots.
Figure~\ref{fig:H2OSilicene} (d) shows that more Si-O chemical bonds
are formed during the AIMD process from 0.3 to 1.0 ps. Therefore,
the chemisorption of water molecules on silicene and the
hydrophilicity of silicene provide potential applications for
silicene-based water molecule sensors and metal-free catalysts for
the oxygen reduction reaction and water dissociation without the
need to introduce dopants or defects.

\section{Conclusion and outlook} \label{sec:Conclusion}

In summary, we demonstrate that the DGDFT method can achieve
extreme-scale HPC KS-DFT calculations on the Sunway TaihuLight
supercomputer. The DGDFT method uses the ALB functions generated
on-the-fly during the SCF iteration to solve the KS equations with
the high precision of plane-wave basis set. In particular, the DGDFT
method adopts a two-level parallelization strategy that makes use of
different types of data distribution, task scheduling, and data
communication schemes, and combines with the feature of master-slave
multi-thread heterogeneous parallelism of SW26010 processor,
resulting in extreme-scale HPC master-slave multi-thread
heterogeneous parallelism on the Sunway TaihuLight supercomputer. We
show that DGDFT can achieve a high parallel efficiency up to 32.3\%
(Speedup as high as 42382.9) by using 8,519,680 processing cores
(131,072 core groups) on the Sunway TaihuLight supercomputer for
studying the electronic structures of two-dimensional (2D) metallic
graphene systems containing tens of thousands of (2,880 and 11,520)
carbon atoms.

For diagonalizing the block-tridiagonal spare DG Hamiltonian in the
DGDFT method, the PEXSI method is more efficient and scalable than
the CheFSI method. The PEXSI method is designed for sparse matrix
operations to take advantage of massively parallel supercomputer
with the high scalability, which can scale up to 100,000
processors.\cite{JCP_143_124110_2015_DGDFT,
PCCP_17_31397_2015_DGDFT} But the PEXSI method is difficult to be
reoptimized and ported on modern heterogeneous supercomputers. In
the future work, we try to optimize and port the parallel
implementation of the PEXSI method on the Sunway TaihuLight
supercomputer.

\vspace{3ex}

\section*{Acknowledgments}

This work is partly supported by National Natural Science Foundation
of China (21688102, 21803066), by Chinese Academy of Sciences
Pioneer Hundred Talents Program (KJ2340000031), by National Key
Research and Development Program of China (2016YFA0200604), Anhui
Initiative in Quantum Information Technologies (AHY090400),
Strategic Priority Research Program of Chinese Academy of Sciences
(XDC01040100),  Research Start-Up Grants (KY2340000094) and Academic
Leading Talents Training Program (KY2340000103) from University of
Science and Technology of China. The authors thank the National
Supercomputing Center in Wuxi, the Supercomputing Center of Chinese
Academy of Sciences, the Supercomputing Center of USTC, and Tianjin,
Shanghai, and Guangzhou Supercomputing Centers for the computational
resources.

\footnotesize{
\bibliography{achemso}
}

\vspace{3ex}

\[
\includegraphics[width=0.5\textwidth]{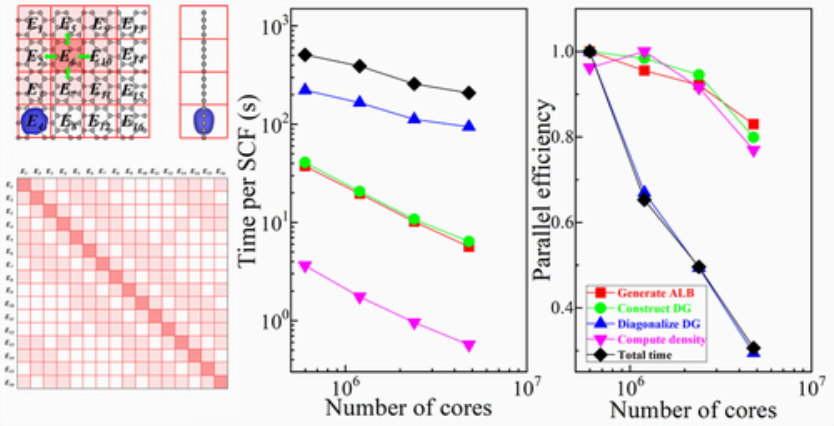}
\]
\centerline{TOC}

\end{document}